# Plasma-state metasurfaces for ultra-intensive field manipulation

**Zi-Yu Chen[1,*], Hao Xu[2], Jiao Jia[2], Yanjie Chen[3], Siyu Chen[2], Yan Zhang[3], Mingxuan Wei[2], Minghao Ma[2], Runze Li[2], Fan Yang[1], Mo Li[2], Guangwei Lu[2], Weijun Zhou[2], Hanmi Mou[2], Zhuofan Zhang[2], Zhida Yang[2], Jian Gao[2,4], Feng liu[2,4], Boyuan Li[2,4], Min Chen[2,4], Liming Chen[2,4], Yongtian Wang[3], Lingling Huang[3,*], Wenchao Yan[2,4,*], Shuang Zhang[5,6,*], Jie Zhang[2,4,7,*]**

[1]Key Laboratory of High Energy Density Physics and Technology (MoE), College of Physics, Sichuan University, Chengdu 610064, China.

[2]State Key Laboratory of Dark Matter Physics, Key Laboratory for Laser Plasmas (MoE), School of Physics and Astronomy, Shanghai Jiao Tong University, Shanghai 200240, China.

[3]Beijing Engineering Research Center of Mixed Reality and Advanced Display, Key Laboratory of Photoelectronic Imaging Technology and System of Ministry of Education of China, School of Optics and Photonics, Beijing Institute of Technology, Beijing 100081, China.

[4]Collaborative Innovation Center of IFSA, Shanghai Jiao Tong University, Shanghai 200240, China.

[5]New Cornerstone Science Laboratory, Department of Physics, University of Hong Kong, Hong Kong, China.

[6]Materials Innovation Institute for Life Sciences and Energy (MILES), HKU-SIRI, Shenzhen 518000, China.

[7]Institute of Physics, Chinese Academy of Sciences, Beijing 100190, China.

These authors contributed equally: Zi-Yu Chen, Hao Xu, Jiao Jia, Yanjie Chen.

[*]e-mail: ziyuch@scu.edu.cn; huanglingling@bit.edu.cn; wenchaoyan@sjtu.edu.cn; shuzhang@hku.hk; jzhang1@sjtu.edu.cn

## Abstract

High-power lasers offer ultrahigh intensities for plasma interactions, but they lack advanced techniques to control the properties of the fields, because no optical elements could withstand their high intensities. The vibrant field of metasurfaces has transformed modern optics by enabling unprecedented control over light at subwavelength through deliberate design. However, metasurfaces have traditionally been limited to solid-state materials and low light intensities. Extending the sophisticated capabilities of metasurfaces from solids into the plasma realm would open new horizons for high-field science. Here, we present a proof-of-concept experimental demonstration of plasma-state metasurfaces (PSMs) via the photonic spin Hall effect and the generation of stable-propagating vortex beams under intense laser irradiation. Time-resolved pump-probe measurements reveal that the functionality of PSMs can persist for several picoseconds, making them suitable for controlling ultra-intense femtosecond lasers, even in state-of-the-art multi-petawatt systems. Harnessing the powerful toolkit of metasurfaces, this approach holds the promise to revolutionize our ability to manipulate the amplitude, phase, polarization, and wavefront of high-power lasers during their pulse duration. It also opens new possibilities for innovative applications in laser-plasma interactions such as compact particle acceleration and novel radiation sources.

## Introduction

Metasurfaces, also known as two-dimensional metamaterials, are structured thin-film devices composed of artificially engineered arrays of subwavelength elements[1-4]. These building blocks, often called "meta-atoms", interact with light in distinctive ways, enabling unprecedented control over light properties, such as its amplitude[5], phase[6], polarization[7], and wavefront[8], based on Huygens' principle. Compared to bulk metamaterials and traditional optical devices, metasurfaces offer significant advantages in terms of their ultrathin profile, lightweight nature, and ease of fabrication. The exceptional and versatile ability of metasurfaces to manipulate light by design has garnered significant attention since their inception. A wide range of innovative applications have been demonstrated, including flat lenses[9,10], polarization multiplexing[11], vortex beam generation[12,13], multicolor holographic imaging[14], optical encryption[15], and dynamically reconfigurable devices[16]. As a result, metasurfaces have become a rapidly evolving field, representing a paradigm shift in optics.

Despite remarkable progress, the metasurfaces explored to date have primarily relied on metallic or dielectric nanostructures, both of which are solid-state materials. Regardless of material composition, solid-state metasurfaces are susceptible to damage, especially when exposed to high-energy or intense light sources. Metallic nanostructures[3] generally have lower damage thresholds, typically around $10^{11}$ W cm$^{-2}$, compared to all-dielectric nanostructures[17], which can withstand intensities one or two orders of magnitude higher. Consequently, most metasurface studies and applications so far have been confined to low laser intensities to prevent device failure[18].

The advent of high-power laser technologies, thanks to invention of the chirped pulse amplification technique[19], has led to a dramatic increase in light intensity in recent decades. State-of-the-art multipetawatt femtosecond lasers can achieve intensities exceeding $10^{23}$ W cm$^{-2}$ [20], while terawatt-class lasers delivering focused intensities above $10^{18}$ W cm$^{-2}$ are routinely operated in many laboratories worldwide[21]. These high-intensity lasers can accelerate electrons to relativistic velocities within a single laser cycle and create extreme material states, i.e., plasmas, which exhibit rich nonlinear physics and can withstand intense fields[22]. Intense laser plasma interactions drive broad important applications, including inertial confinement fusion[23], laboratory astrophysics[24], compact particle acceleration[25,26], and secondary radiation sources[27,28]. However, the current techniques for manipulating intense light are still highly limited. Expanding the sophisticated light-shaping capabilities offered by metasurfaces to high-field science would represent a significant advancement, opening new avenues for controlling intense light and exploring related applications.

When irradiated by an intense laser pulse, the surface of solid-state metasurfaces ionize to form plasmas (Fig. 1a). Due to the ultrashort duration of these laser pulses (typically less than 100 fs), the ions of the surface material hardly have time to expand (expansion speed is limited by the slow ion sound speed), thus preserving the shape and density of the original solid target (Fig. 1b). Therefore, plasma-state metasurfaces (PSMs) can be generated by irradiating pre-structured solid-metasurfaces with intense ultrashort lasers. Pancharatnam-Berry (PB) metasurfaces rely on geometric phase effects, which are inherently linked to the structure of the metasurface, rather than material composition or state. This underlying mechanism of geometric phase suggests that PB metasurfaces could potentially function under intense light conditions, provided their structural integrity is preserved. Yet, the impact of ionized free electrons on metasurface functionality is unclear. Although ion dynamics primarily shape plasma structure, electron response dictates the reflection of electromagnetic waves.

Here, we present PSMs composed of fully ionized plasmas and experimentally validate their ability to function effectively when interacting with intense laser fields during the laser duration. As a proof of concept, we utilize the photonic spin Hall effect in linear phase-gradient PB metasurfaces as a representative example[29,30] (Fig. 1c). We experimentally observe the photonic spin Hall effect for an intense light beam for the first time in a plasma medium, which offers new opportunities for strong-field applications. Time-resolved pump-probe measurements show that the functional lifetime of the PSMs lasts for several picoseconds. The superior capabilities of PSMs compared to conventional optics in high-intensity light manipulation are further showcased by the generation of intense stable-propagating optical vortices, surpassing the limits of conventional optics. Additionally, we perform three-dimensional (3D) particle-in-cell (PIC) simulations to demonstrate the effectiveness of PSMs in the relativistic regime and highlight the advancements they enable in high-filed science applications, such as efficient electron and proton acceleration. Going beyond conventional solid-state materials, this concept paves the way for a new era of metasurfaces, holding significant potential for advancing research in high-field science.

## Results

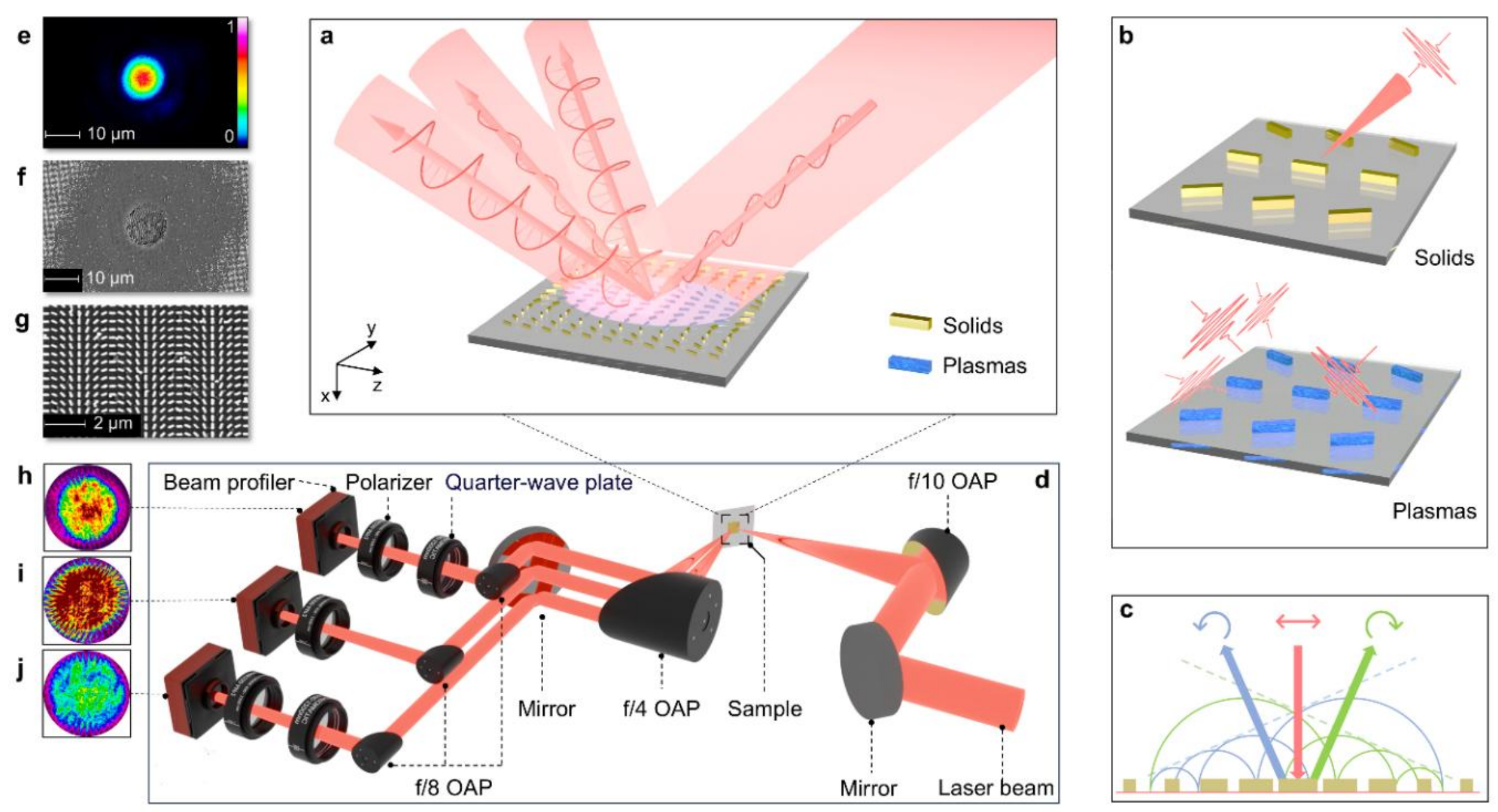

**Fig. 1. Concept of plasma metasurfaces and experimental setup for photonic spin Hall effect. a,** A schematic illustration of the photonic spin Hall effect in Pancharatnam–Berry (PB) geometric-phase metasurfaces composed of plasmas. When irradiated with a linearly polarized (LP) laser beam, the outgoing light splits into three distinct beams: two anomalously reflected circularly polarized (CP) beams

with opposite handedness and one normally reflected LP beam. **b,** Generation of plasma-state metasurfaces by irradiating initially solid metasurfaces with intense femtosecond laser pulses. The plasmas hardly expand due to the ultrashort pulse duration, preserving the shape and density of the original solid target. **c,** Working principle of PB phase metasurfaces inducing the photonic spin Hall effect. CP lasers with opposite handedness experience opposite phase gradients, resulting in different wavefronts. **d**, A schematic layout of the experimental setup to measure the photonic spin Hall effect. **e,** CCD image of the laser focal spot measured by a laser beam profiler. **f, g,** Scanning electron microscopy (SEM) images of the metasurfaces sample after and before intense laser irradiation, respectively. **h-j,** Typical CCD images recorded by beam profilers for the three reflected beams with different polarization states, respectively.

**Proof of concept through the photonic spin Hall effect**

An anisotropic cuboid is selected as the meta-atom for the PB phase metasurfaces. When this meta-atom is illuminated by a circularly polarized (CP) incident light with electric fields in the *y*-*z* plane of the metasurfaces, expressed as $\begin{pmatrix} E_y^{\mathrm{in}} \\ E_z^{\mathrm{in}} \end{pmatrix} = \frac{1}{\sqrt{2}} \begin{pmatrix} 1 \\ i\sigma \end{pmatrix}$, where $\sigma = \pm 1$ denotes the CP light handedness, the electric fields of the outgoing light can be expressed as[13]

$$\begin{pmatrix} E_y^{\mathrm{out}} \\ E_z^{\mathrm{out}} \end{pmatrix} = \frac{1}{2\sqrt{2}} \left[ (r_{\mathrm{u}} + r_{\mathrm{v}}) \begin{pmatrix} 1 \\ i\sigma \end{pmatrix} + (r_{\mathrm{u}} - r_{\mathrm{v}}) e^{2i\sigma\theta} \begin{pmatrix} 1 \\ -i\sigma \end{pmatrix} \right], \quad (1)$$

where $r_{\mathrm{u}}$ and $r_{\mathrm{v}}$ are the reflection coefficients for two orthogonal linearly polarized (LP) light along the two primary axes of the meta-atom, respectively, and $\theta$ denotes the rotation angle of the meta-atom from the *x*-axis. Importantly, the outgoing light not only includes CP light with the original chirality, but also generates CP light with opposite handedness accompanied by an additional geometric phase shift of $2\sigma\theta$.

By arranging multiple cuboids with orientation angles ranging from 0 to $\pi$ in a unit cell, we can achieve a phase change spanning $2\pi$. Subsequently, the periodic repetition of this unit cell in the $y$-direction imposes a linear geometric phase gradient $\xi = d\Phi(y)/dy$ on the light, where $\Phi(y)$ is phase shift. According to the generalized Snell's law, the in-plane incident and reflected wave vectors parallel to the target surface satisfy[1,2]

$$k_{\parallel}^{\mathrm{r}} = k_{\parallel}^{\mathrm{in}} + \xi = k_{\parallel}^{\mathrm{in}} + \sigma \frac{\Delta\alpha}{S_y}, \tag{2}$$

where $\Delta\alpha$ and $S_y$ are the angle difference and spacing along the $y$-axis between adjacent meta-atoms, respectively. As a consequence, the geometric phase term not only reverses the light's handedness but also causes a shift in the reflection angle. Notably, left-handed CP (LCP) light and right-handed CP (RCP) light experience reflection angle shifts in opposite directions.

When a LP light beam, which can be decomposed into two CP components with opposite handedness, is incident on the metasurface, the outgoing light is split into three distinct beams. Two of these beams (LCP and RCP light) are anomalously reflected due to the geometric phase gradient, while the third beam (LP light) unaffected by the PB phase is normally reflected. This phenomenon demonstrates the photonic spin Hall effect. Similarly, vortex beams carrying orbital angular momentum can be generated by metasurfaces with azimuthal phase singularities.

As indicated by equation (1), anisotropic reflection coefficients, with distinct values for $r_{\mathrm{u}}$ and $r_{\mathrm{v}}$, are essential for generating geometric effects. This implies that metasurfaces composed of plasma-state materials could also be effective, given the common occurrence of anisotropic light responses in laser-plasma interactions.

The experiment setup to demonstrate PSMs and to measure the photonic spin Hall effect is presented in Fig. 1d (Methods). We focus a laser pulse of linear polarization with wavelength $\lambda_0 = 800$ nm onto a metasurface down to a spot size of about 7.9 μm (full width at half maximum, FWHM; Fig. 1e). The linear-phase-gradient metasurface consists of gold nanocuboid arrays grown on a glass substrate (see Methods for details). Figure 1f, g shows scanning electron microscope (SEM) images of the sample after and before the creation of plasmas, respectively (see Supplementary Fig. S1 for magnified SEM images). We measure the three reflected beams with CCD beam profilers, the typical images of which are shown in Fig. 1h-j, respectively.

In the first set of experiments, we position the target away from the laser focus to maintain the on-target intensity below the damage threshold of the metasurface so as to provide a stable reference. The laser pulse energy is 10 μJ (with an estimated on-target energy of about 0.2 μJ). The laser spot size on target is about 138 μm (FWHM), resulting an estimated peak intensity of about $2.2 \times 10^{10}$ W $cm^{-2}$. Three reflected beams are detected, as illustrated in Fig. 2a. The upper and lower panels correspond to the anomalously reflected beams, while the middle panel represents the specularly reflected beam. This observation is characteristic of the photonic spin Hall effect, indicating the normal functioning of the metasurface. Variations in image intensity are attributed to differences in experimental setup adjustments and attenuation for the three beams. By repeatedly irradiating the same target spot for multiple shots under identical conditions, we consistently observe similar three-beam patterns. The measured pulse energy, normalized to the maximum value of each group, is shown in Fig. 2b. These results remain largely unchanged across multiple shots, suggesting that the nanostructures are not damaged by the laser pulses and that the metasurface maintains its solid-state configuration.

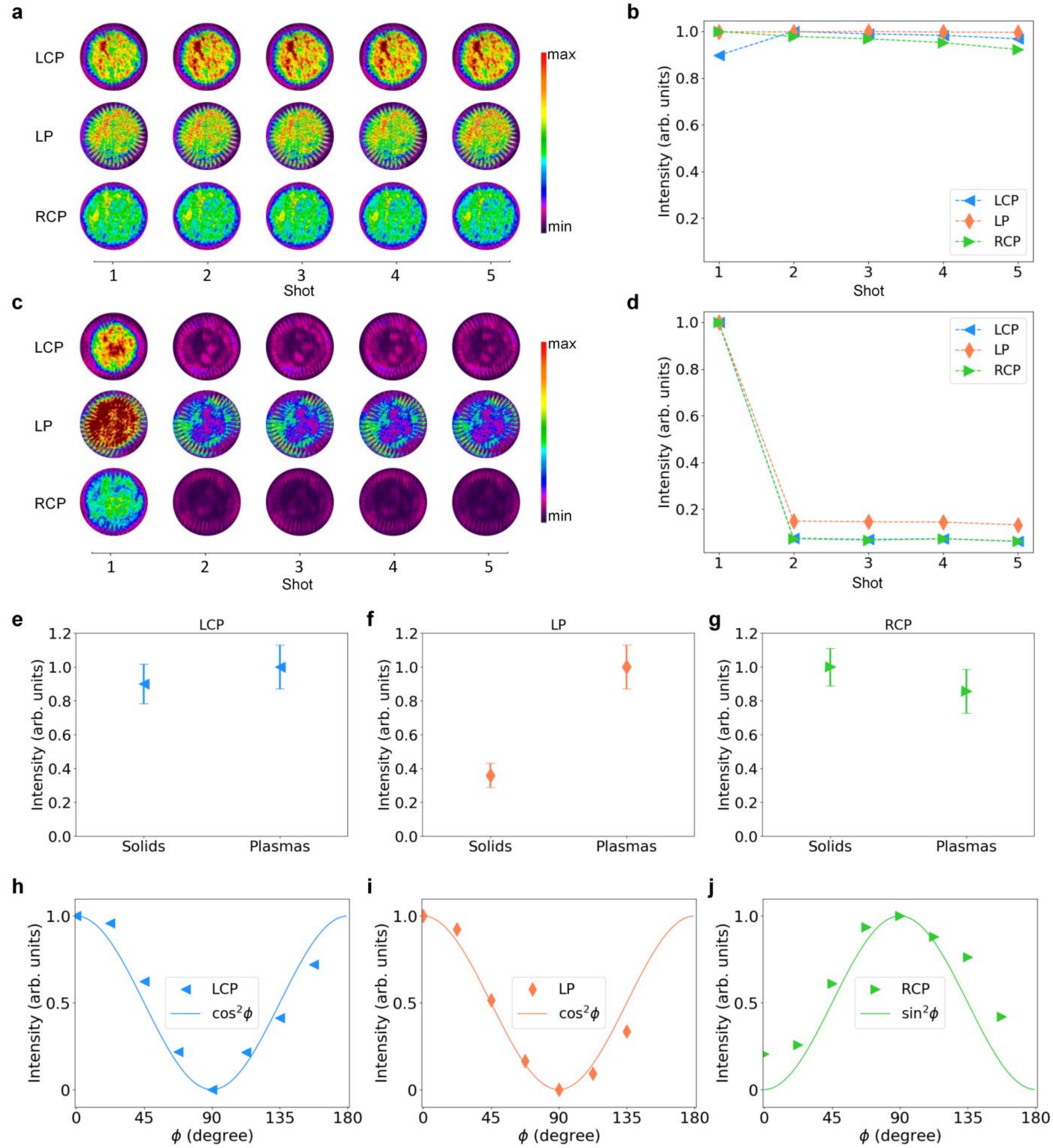

**Fig. 2. Experimental demonstration of photonic spin Hall effect from plasma-state metasurfaces**. **a,** CCD images displaying the intensity profiles of the three reflected light beams from solid-state metasurfaces: LCP (upper panel, left-handed circularly polarized, anomalous reflection), RCP (lower panel, right-handed circularly polarized, anomalous reflection), and LP (middle panel, linearly polarized,

normal reflection). The metasurfaces are irradiated with LP laser beams, each with an intensity of approximately $2.2 \times 10^{10}$ W cm$^{-2}$, at the same target spot for multiple shots. **b,** Integrated light intensity vs. shot number for the three reflected beams corresponding to **a**. **c,** CCD images of the three reflected light beams from plasma-state metasurfaces at an incident laser intensity of $4.5 \times 10^{12}$ W cm$^{-2}$. Each shot is positioned at the same target spot. **d,** Variation of integrated light intensity for different shots corresponding to **c**. Intensities in **b** and **d** are normalized to the maximum value in each group. **e-g**, Comparison of reflected light intensity between solid-state and plasma-state metasurfaces for (**e**) LCP, (**f**) LP, and (**g**) RCP reflected light beams. The first shots at laser intensities of $2.2 \times 10^{10}$ W cm$^{-2}$ and $4.5 \times 10^{12}$ W cm$^{-2}$ are taken for the solid-state and plasma-state metasurfaces, respectively. **h-j,** Measured light intensity as a function of $\phi$ (angle between the polarization direction of the light and the transmission axis of the linear polarizer) from plasma-state metasurfaces for (**h**) LCP, (**i**) LP, and (**j**) RCP reflected light beams. Solid curves represent calculations based on Malus's law. For each shot, the laser intensity is maintained at $4.5 \times 10^{12}$ W cm$^{-2}$, and a fresh target spot is provided.

In the second set of experiments, the target is positioned at the laser focus to ensure a high peak intensity on the metasurfaces. While the laser pulse energy remains unchanged, the spot size on the target is significantly reduced to about 7.9 μm (FWHM). This results in an estimated peak intensity of $4.5 \times 10^{12}$ W cm$^{-2}$. Note that our metasurface is fabricated from gold, a metallic material. The ionization and ablation threshold for thin gold films and nanostructures under femtosecond laser irradiation is significantly lower than that of dielectrics. Even for dielectric metasurfaces, considering the nanostructures, unless they are specially designed for high power, the damage threshold is typically far lower than $10^{13}$ W cm$^{-2}$. For examples, the experimentally measured damage thresholds for all-dielectric silicon metasurfaces[18] and wide-bandgap semiconductor gallium phosphide metasurfaces[31] are on the order of $10^{11}$ W cm$^{-2}$. Consequently, the damage threshold for our gold nanostructures is well below $4.5 \times 10^{12}$ W cm-

$^{2}$. This experimental intensity is sufficient to transition the gold nanostructured metasurfaces into a fully ionized, high-density plasma state. For the initial shot, three reflected beams similar to those observed at lower intensities are detected (Fig. 2c). However, subsequent shots reveal distinct image features. A crater surrounded by a halo of reflected light becomes clearly visible in all these shots. The normalized measured energy exhibits a dramatic decrease from the first to second shot and then remains relatively constant (Fig. 2d). An examination of the laser focal spot intensity distribution reveals a Gaussian profile with a bright central spot (Fig. 1e). The intensity within this central spot is high enough to generate plasmas. Despite the plasma state, the structure pattern of the metasurfaces remains essentially unchanged during the laser-plasma interactions due to the ultrashort duration of the laser pulse, which limits plasma expansion. Consequently, we can observe signals from the entire area of the laser spot. After the first shot, however, the heated and expanded plasmas partially destroy the metasurfaces nanostructures within the central spot (see Fig. 1f and Supplementary Fig. S1), creating a dark hole in the center, particularly evident in the images of the two anomalously reflected beams (Fig. 2c). The residual reflection within the circle is likely attributed to stray light or remnant structures. The structure-destroyed area observed outside the laser spot circle in Fig. 1f can be attributed to an ablation process that occurs at a much later time after the laser reflection, as indicated by the time-resolved pump-probe measurements of the plasma metasurface lifetime discussed below. In contrast, the lower intensity in the outer regions of the focal spot allows the metasurfaces structures to survive (Fig. 1g), resulting in a persistent halo of reflection in the follow-up shots (Fig. 2c). The image contrast between the two sets of experiments clearly indicates that the first shot of the second set of experiments produces plasmas in the central spot.

A comparison of the reflected light intensity from samples in solid and plasma states (Fig. 2e-g) reveals that the normally reflected beam (denoted by LP) from the plasma-state metasurfaces exhibits a higher intensity compared to the solid-state metasurfaces. This enhancement is possibly due to the plasma mirror effect, as the plasma is too dense to permit laser transmission. The intensities of the anomalously reflected beams (denoted by LCP and RCP) are comparable in both plasma and solid states, ruling out the possibility that the signal measured in the first shot of the second set of experiments originated from solid-state metasurfaces induced by the low-intensity leading edge of the laser pulse, because the leading edge comprises only a small fraction of the total laser pulse energy. Therefore, the measured signal in the first shot of the second set of experiments should be attributed to the main pulse reflection from the metasurfaces in the plasma state.

To confirm that the three beams generated in the plasma state are indeed a result of the photonic spin Hall effect, we measure the polarization state of each beam. The laser intensity is maintained at $4.5 \times 10^{12}$ W cm$^{-2}$, and each shot is directed to a fresh spot on the target. The anomalously reflected beams are analyzed using a quarter-wave plate followed by a linear polarizer, while the normally reflected beam is measured with a single linear polarizer. The quarter wave plate is employed to convert CP light into LP light, which can then be analysed using the linear polarizer. The measured intensity of the transmitted light as the linear polarizer is rotated is shown in Fig. 2h-j. The results are in good agreement with the Malus's law, $I_{\mathrm{t}} = I_0 \cos^2\theta$, where $I_0$ and $I_{\mathrm{t}}$ are the light intensities before and after the linear polarizer, respectively, and $\theta$ is the angle between the polarization axis of the incident light and the transmission axis of the polarizer. This verifies that the two anomalously reflected beams are CP light, while the normally reflected beam is LP light. Furthermore, the observed π/2 phase shift between the two anomalously reflected beams indicates that they are CP light with opposite helicity, thus confirming the photonic spin Hall effect exhibited by the plasma-state metasurfaces. This work constitutes the first successful proof-of-principle demonstration of intense light field manipulation using a metasurface in its plasma state. It is also the first time the photonic spin Hall effect has been experimentally observed and controlled in a plasma-state material and under strong laser field conditions. Demonstrating that this fundamental geometric phase effect persists and can be engineered within a high-density plasma represents a significant and novel advancement in the field of plasma photonics.

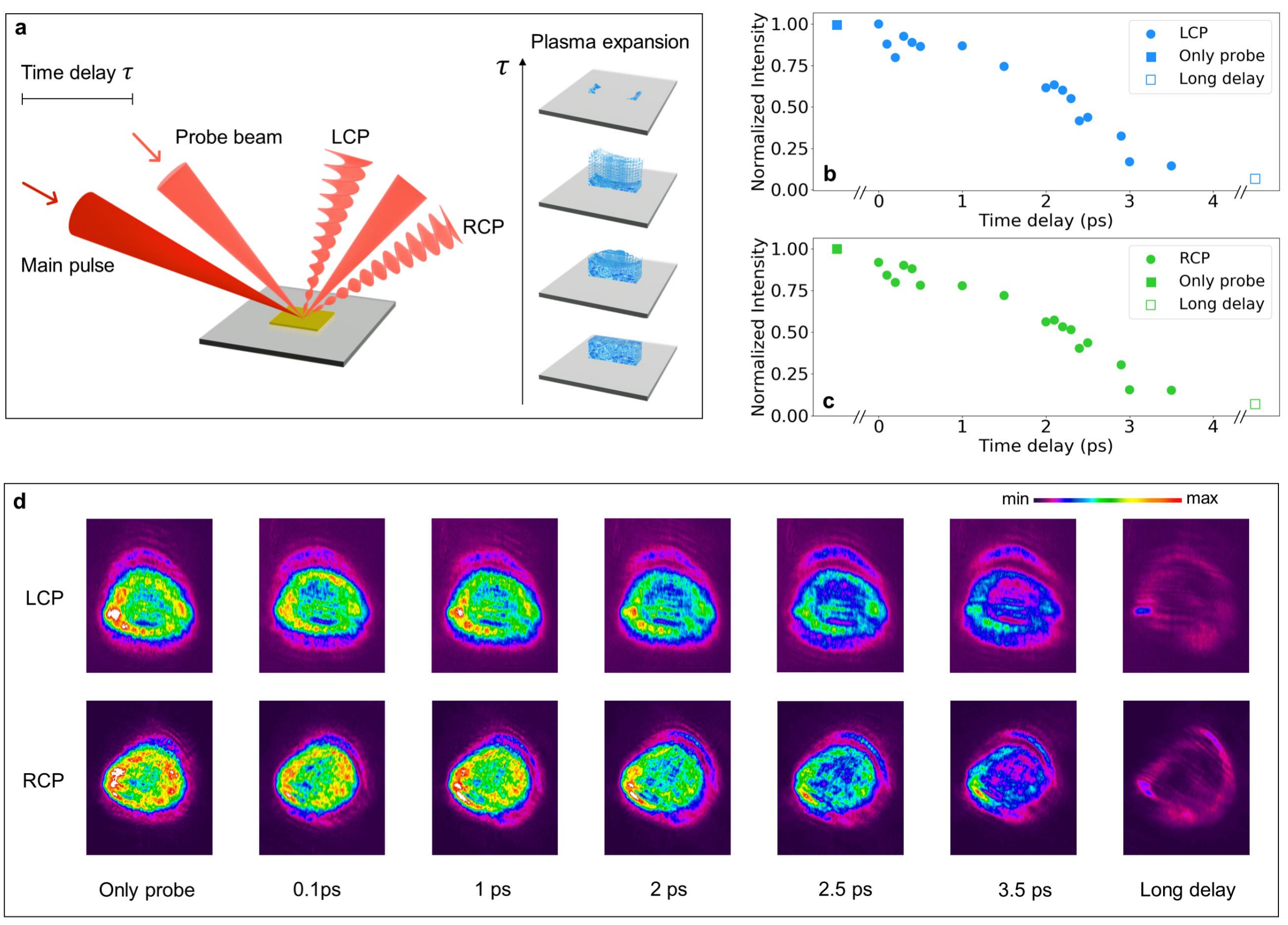

**Fig. 3. Time-resolved pump-probe measurement of plasma metasurface lifetime. a,** A schematic diagram of the pump-probe experiment. An intense main pulse (pump beam with intensity of $\sim 1.7 \times 10^{15}$ W cm$^{-2}$) irradiates the target, transforming it into a plasma state. As the plasma expands over time, it gradually smears out the metasurface structures. A second, weaker pulse (probe beam with intensity of $\sim 2.5 \times 10^{14}$ W cm$^{-2}$) with linear polarization, split from the main pulse, is used to interrogate the functionality of the plasma metasurfaces at various time delays after the initial irradiation. The functionality is assessed by measuring the LCP and RCP reflected light intensities induced by the photonic spin Hall effect. **b, c,** The experimentally measured LCP and RCP reflected light intensities, respectively, as a function of time delay between the pump and probe pulses. The intensities are normalized to those obtained using the probe beam alone. For each delay time measurement, a fresh target area is used. The long delay time corresponds to several minutes. **d,** CCD images of the LCP and RCP reflected light intensity profiles as a function of probe delay.

## Operational lifetime of PSMs

To effectively manipulate high-intensity light using PSMs, it is crucial to maintain the metasurface structures for a duration longer than the light pulse before they are destroyed by plasma hydrodynamic expansion. To determine the functional lifetime of these plasma metasurfaces, we conduct pump-probe experiments in a separate experimental campaign. This run requires a completely different optical layout optimized for the demanding technical challenge of time-resolved experiments to measure the ultrafast plasma dynamics. The schematic diagram of the experiments is illustrated in Fig. 3a. A weaker probe beam with linear polarization (intensity $\sim 2.5 \times 10^{14}$ W cm$^{-2}$), split from the main laser pulse (intensity $\sim 1.7 \times 10^{15}$ W cm$^{-2}$), is introduced with an adjustable time delay (Methods and Supplementary Fig. S2). Figure 3b, c shows the time-resolved reflected intensities of the probe beam as a function of the time delay between the main and probe pulses. The presence of both LCP and RCP probe beams, resulting from the photonic spin Hall effect, indicates the sustained functionality of the PSM. Both signals exhibit a similar decay trend over time. Figure 3d shows the CCD images of the LCP and RCP reflected light intensity profiles as a function of the time delay, which exhibit slight distortion because, during this round of experiments, our primary focus is to compare the energy evolution of the split beam. The beam spot shape is not specifically optimized. Nevertheless, this minor beam deformation does not impact the validity of our conclusions. Our findings reveal that the PSM maintains its functionality for about 3 ps, consistent with previous experimental results on plasma expansion speed[32]. Particularly, within a temporal window of 1 ps after the main laser irradiation, the probe beams retain approximately 90% of the intensity observed with solid-state metasurfaces. The lifetime of PSMs is intrinsically limited by the ion expansion within the plasma, making it challenging to extend their operation to longer or repetitive pulses. Nevertheless, the significance of our work lies in the control of ultraintense lasers, which are typically ultrashort femtosecond pulses and under single-shot conditions. An approximately 3 ps lifetime is more than sufficient for manipulating such femtosecond-scale laser fields. In comparison, solid-state metasurfaces may be better suited for applications requiring long pulses or repetitive operation; however, their utility is limited by optical damage thresholds that cap laser intensity. Therefore, advancing both PSMs (for extreme-intensity, single-shot regimes) and high-damage-threshold solid-state metasurfaces (for repetitive, moderate-intensity applications) represents two important and complementary pathways for the future development of intense light control platforms.

## PSMs in the relativistic regime

The concept of PSMs for high-field manipulation is demonstrated by our proof-of-principle experiments. The next critical step is to operate at a higher intensity, exceeding $10^{18}$ W cm$^{-2}$. This intensity regime is particularly interesting because it involves significant nonlinear and relativistic effects. To confirm feasibility, we perform 3D PIC simulations that included these extreme nonlinear and relativistic effects using the VLPL code[33] (see Methods for detailed simulation parameters). The full experimental demonstration in the relativistic regime will be addressed in future work and published separately.

Initially, we numerically calculate the optical response of an individual cuboid meta-atom composed of fully-ionized plasmas. The LP laser intensity is set at $2.0 \times 10^{18}$ W cm$^{-2}$. At this intensity, the plasma is overdense with a uniform density of $n_{\mathrm{e}} = 400 n_{\mathrm{c}}$ (where $n_{\mathrm{c}} \approx 1.7 \times 10^{21}$ cm$^{-3}$ is the critical plasma density) and therefore opaque to the laser. We obtain the ratio between the complex amplitude of the reflected electric field along the two primary axes of the meta-atom as $r_{\mathrm{u}}/r_{\mathrm{v}} \approx 0.419 e^{2.26i}$. Subsequently, the efficiency of the PB phase is estimated as $\eta = \left|r_{\mathrm{u}} - r_{\mathrm{v}}\right|^2 / \left(\left|r_{\mathrm{u}} - r_{\mathrm{v}}\right|^2 + \left|r_{\mathrm{u}} + r_{\mathrm{v}}\right|^2\right)^2 \approx 73\%$, demonstrating the potential of PB phase plasmas metasurfaces in the relativistic regime.

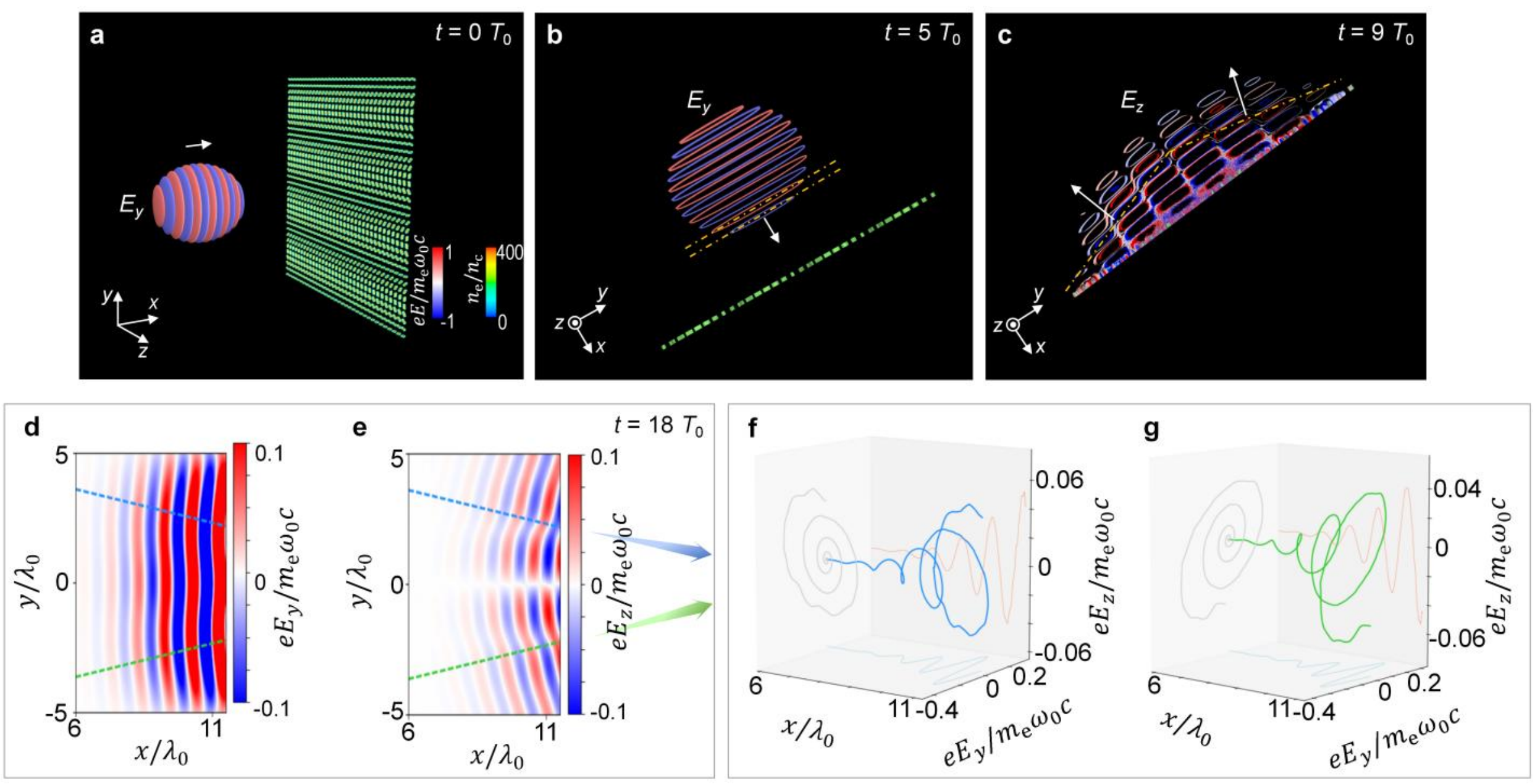

**Fig. 4. 3D PIC simulations of photonic spin Hall effect from plasma metasurfaces in the relativistic regime. a,** 3D isosurfaces of the incident laser filed $E_y$ (blue-red color scale) and plasma metasurface density $n_{\mathrm{e}}$ (green color scale) at the initial time $t = 0$. A linearly polarized (LP) laser pulse with normalized amplitude $a_0 = 1.5$ (above the relativistic threshold of $a_0 = 1$), corresponding to an intensity of $4.8 \times 10^{18}$ W cm$^{-2}$, is incident normally on a target with density $n_{\mathrm{e}} = 400n_{\mathrm{c}}$. The PB phase metasurface comprises four unit cells in the *y*-direction, each containing eight plasma cuboids with varying orientations to induce a phase gradient in the *y*-direction. The white arrow indicates the wave propagation direction. **b,** Sliced isosurfaces of $E_y$ and $n_{\mathrm{e}}$ at a later time $t = 5T_0$ ($T_0$ is the laser period), which is before the laser-plasma interaction. The yellow dash dotted lines indicate the incident wavefronts, showing normal propagation towards the target. **c,** Sliced isosurfaces of the reflected field $E_z$ and $n_{\mathrm{e}}$ at $t = 9T_0$, which is during the laser-plasma interaction. The yellow dash dotted lines indicate the reflected wavefronts, revealing anomalous deflection into two beams. **d, e,** 2D electric filed profiles in the *x*-*y* plane at $t = 18T_0$ after the reflection for the (**d**) $E_y$ and (**e**) $E_z$ components of the reflected pulse, respectively. Blue and green dashed lines indicate the anomalous reflection directions induced by the metasurfaces. **f, g,**

Reconstructed 3D electric field vector images along the anomalous reflection directions in the (**f**) upper and (**g**) lower half spaces, respectively.

We then simulate the photonic spin Hall effect from the PSM in the relativistic regime with LP incident laser pulses (Fig. 4). To observe this effect in PIC simulations, we analyze the electric fields to identify the spatial separation of the LCP and RCP components, as photon spin is directly manifested in the circular polarization of light. To clearly isolate PSM's wavefront engineering effects, the incident laser pulse is initially linearly polarized along the $y$-axis, with its electric field oscillating solely in the $E_y$ component. The laser propagates along the $x$-direction with normal incidence. We note that PSMs also function at oblique incidence in the relativistic regime, as demonstrated in the following by 3D PIC simulations of relativistic vortex beam generation (Fig. 5) and particle acceleration enhanced by surface plasmon coupling (Supplementary note 1). A snapshot of 3D isosurfaces of the incident laser field and PSM density at the initial time $t = 0$ is shown in Fig. 4a. Sliced isosurfaces of $E_y$ and $n_e$ at a later time $t = 5T_0$ (before the laser-plasma interaction; $T_0$ is the laser period) are presented in Fig. 4b, exhibiting predominantly plane wavefronts of the incident beam with only the $E_y$ component. Upon reflection from the plasma metasurface, the spin-dependent PB phases cause the reflected light to split into two spatially separated, CP pulses of opposite handedness. Each of these resultant pulses now possesses both $E_y$ and $E_z$ electric field components (Fig. 4c). The spatial modulation of the wavefront, as evidenced by the dash dotted lines in Fig. 4c, indicates anomalous reflection into two beams. Figure 4d, e display the 2D electric field profiles in the $x$-$y$ plane at $t = 18T_0$ (after reflection from the metasurfaces) for the $E_y$ (Fig. 4d) and $E_z$ (Fig. 4e) components of the reflected pulse, respectively. The blue and green dashed lines mark the directions of anomalous reflection, calculated using the generalized law of refraction[1]. By analyzing the relative amplitude and phase of the $E_y$ and $E_z$ components, we can construct a 3D profile of the total reflected electric field. The reconstructed field vectors along the anomalous reflection directions in the two half spaces are shown in Fig. 4f, g, respectively. This visualization clearly confirms the opposite handedness of the two pulses. The amplitude of the $E_y$ component is larger than that of the $E_z$ component in the anomalously reflected beam due to the presence of the ordinary $E_y$ field, which is not affected by the geometric phase. The three beams overlap due to the limited propagation distance from the metasurfaces, which is chosen to save computation time. Nevertheless, the two anomalously reflected beams with opposite handedness clearly demonstrate the photonic spin Hall effect. Similar simulation results incorporating preplasmas effects demonstrate the robustness of the PSM in the relativistic regime (Supplementary Fig. S3). Note that free ionized electrons, heated by the relativistic laser to high energies, are taken into account in the simulations. The electron density distribution reveals that the primary structure of the metasurface is well maintained (Supplementary Fig. S4). This is because the strong electrostatic force between ions and electrons in high-density plasmas attracts most electrons to the ions. Thus, PSMs composed of fully-ionized plasmas can retain their functionality even in high-field regimes where electron dynamics become relativistic.

## Application of PSMs on intense stable propagating vortex beams generation

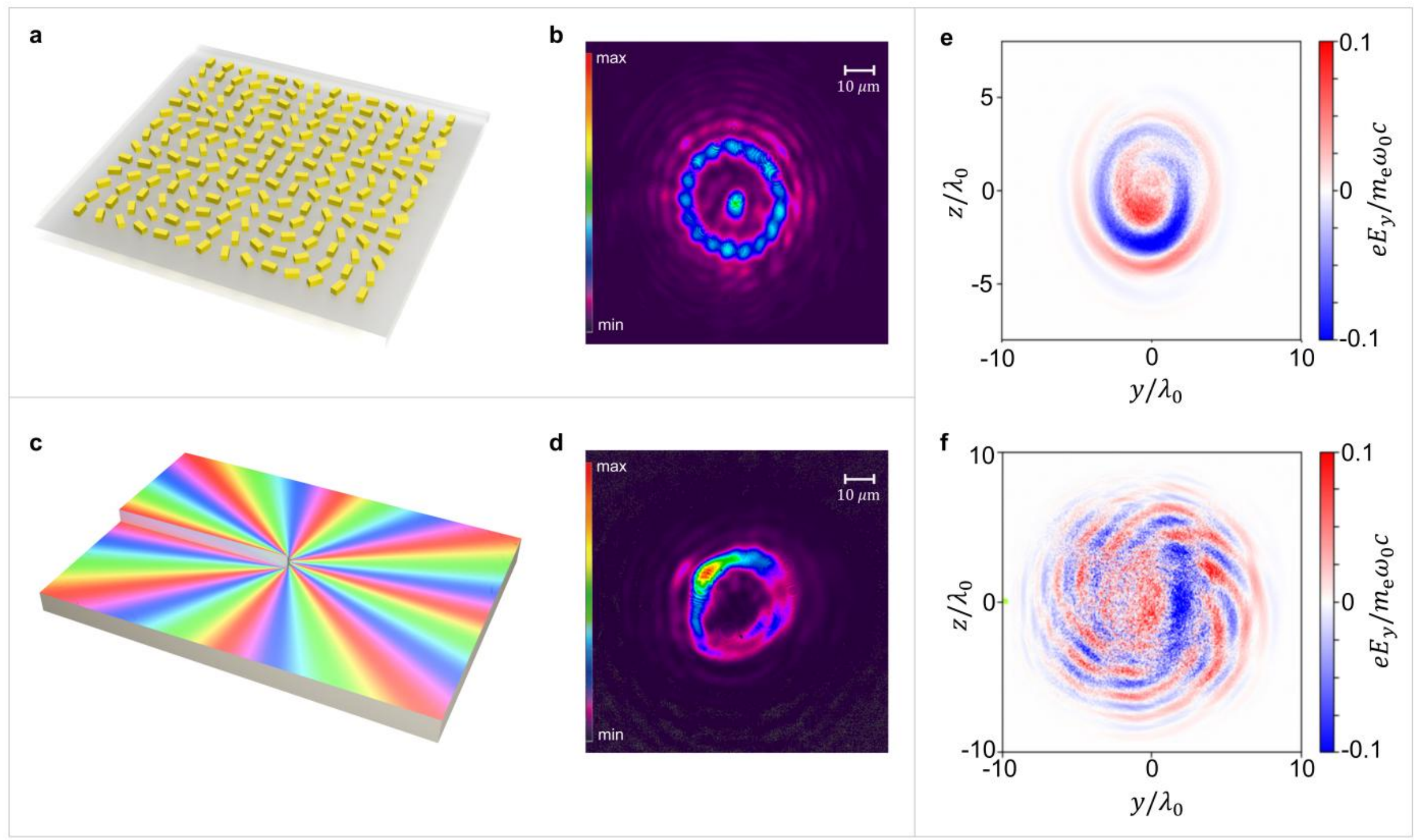

**Fig. 5. Optical vortex generation by plasma metasurfaces and spiral phase mirrors. a,** Schematic of the designed PB metasurfaces sample with a topological charge of $l = 8$ and size of $500$ μm × $500$ μm. **b,** Experimentally measured intensity profile of optical vortices generated by the plasma PB metasurfaces. **c,** Schematic of the spiral phase mirror sample with a size of 70 mm × 50 mm designed for wavelength of 800 nm and topological charge of $l = 8$. **d**, Experimentally measured intensity pattern of vortex beam generated by the spiral phase mirror. The femtosecond laser pulse is centered at 800 nm with a broad bandwidth (770 to 830 nm). **e, f,** 3D simulation results in the relativistic regime for the interference pattern of the electric fields $E_y$ between the PB-phase-affected vortex pulse and the unaffected Gaussian pulse generated by plasma metasurfaces with a topological charge of (e) $l = 1$ and (f) $l = 8$, respectively.

Having demonstrated the concept of PSMs, we provide more compelling experimental evidence demonstrating how this approach enables a significant advance in performance or manipulation capabilities in high-field science applications. As a showcase, we conduct experiments to generate intense stable propagating optical vortices from a PSM, which exhibits significant advantages over conventional optics. High-intensity optical vortices are highly sought after for numerous laser-plasma applications but pose significant technical challenges in their generation[34-37]. Traditional methods for creating these special light beams often rely on bulky devices called spiral phase plates or mirrors. These plates or mirrors are designed for a specific wavelength of light, which can be a problem when using ultrashort femtosecond lasers due to their broad spectral widths. Additionally, the specially designed spiral phase mirrors can be highly effort consuming, especially for large multi-petawatt lasers. In contrast, plasma metasurfaces offer broadband operation, crucial for generating vortex beam with integer topological charge, which is an eigenstate of propagating light. We design PB PSMs with a topological charge of $l = 8$ (Fig. 5a). The experimental setup to measure the vortex beam is detailed in Methods and Supplementary Fig. S5. Figure 5b presents the measured reflected light intensity from the PSM at an intensity of $1.4 \times 10^{14}$ W cm$^{-2}$. The annular intensity distribution, characteristic of a vortex beam, confirms its generation via the PB phase effect. The appearance of approximately 16 lobes in the intensity distribution is a signature of the coherent superposition of the main vortex beam with a small amount of an oppositely charged vortex mode. This minor, counter-rotating component arises from slight imperfections or unwanted polarization effects in the setup[38]. The interference between $+l$ and $-l$ modes is well-known to produce an intensity profile with $2l$ angular lobes. This phenomenon is extensively documented in the literature on vortex beam generation from metasurfaces[39]. Critically, even a very minor component with the opposite topological charge (a few percent or less) is sufficient to form this $2l$ lobed structure[38,39]. To experimentally verify the intended topological charge, we perform a standard interferometry test. The interference pattern between the metasurface-generated vortex beam and the PB phase-unaffected Gaussian reference beam exhibits 8 distinct spiral lobes (Supplementary Fig. S6), confirming the topological charge of the dominant generated beam being $l = 8$. A quarter waveplate and a linear polarizer are used to filter out the PB-phase unaffected Gaussian beam with opposite circular polarization. Note that such vortex beams can also be deflected away from the Gaussian beam through the design of a phase gradient within the PSMs. Interestingly, a central intensity spot emerges, which is a distinct characteristic of the plasma-state metasurfaces and is not observed in the low-intensity solid-state metasurface counterpart (Supplementary Fig. S6). This spot is likely caused by the polarization state change of the central reflected Gaussian light due to the plasma dynamics in the central plasma region. As demonstrated by Chen et al.[40], when a laser pulse is obliquely incident onto a solid-density plasma surface, the reflection process can induce a phase shift between different polarization components. This altered polarization state means this part of light reflected from the central plasma region can no longer be perfectly filtered by the downstream polarization optics, resulting in the observed bright spot. For comparison, we also generate vortex beams using a conventional spiral phase mirror designed for 800 nm and a topological charge of $l = 8$ (Fig. 5c). The resulting intensity profile exhibits a nonuniform C-shape (Fig. 5d), indicating that most energy carries fractional topological charges, attributed to the laser pulse's broad bandwidth (770 to 830 nm). Fractional vortex beams, being a superposition of eigenmodes, can become distorted or lose their coherence during propagation. This, in turn, may result in the failure of applications such as positron acceleration due to unstable hollow-channel accelerating structures[41]. The extraordinary abilities enabled by PSMs to generate intense stable-propagating optical vortices and to realize complex functionalities, such as arrays of vortices with spatially varying orbital angular momenta[12], are essential for advanced high-field applications ranging from efficient particle acceleration[34,37,41] to high harmonic[35] and magnetic field generation[36] using vortex beams.

Figure 5e, f displays the 3D PIC simulation results for the vortex beam generation at relativistic laser intensities from a PSM designed with a topological charge of $l = 1$ and $l = 8$, respectively (see Supplementary Fig. S7 for the interaction scheme and plasma-targets design). We consider both normal and oblique incidence cases, and find that the reflected beam's structure is not affected. The interference pattern of the electric fields between the PB-phase-generated vortex beam and the unaffected Gaussian pulse in Fig. 5e, f reveals topological charges of $l = 1$ and $l = 8$, respectively, thereby confirming the effectiveness of the PSM in generating stable propagating optical vortices with integer topological charges considering relativistic effects. Additionally, we perform simulations for relativistic laser-driven electron and proton accelerations using PSMs as highly-efficient plasma-targets to further showcase their great potential in high-field science applications (Supplementary note 1). Our results show that PSM targets are capable of producing significantly higher particle yields and cutoff energies compared to conventional plane targets (Supplementary Fig. S8 and S9), because PSMs can be designed with a large phase gradient to enable very efficient laser energy coupling to surface plasmons[2]. These findings have crucial implications for fields such as hadron cancer therapy.

## Discussion

As with other single-use plasma optics, PSMs are destroyed after irradiation. While this is common for high-power plasma optics, it may restrict their use in certain contexts. By positioning the metasurfaces upstream of the laser focus, the non-destroyed device can be utilized for multiple shots. This multi-shot capability may benefit the use of metasurfaces for high-power laser manipulation. However, transitioning to plasma-state metasurfaces offers several key advantages. First, PSM devices overcome the damage threshold limitations of traditional solid-state metasurfaces, allowing for laser intensities orders of magnitude higher. The multi-shot benefit of solid-state metasurfaces comes at the expense of maximum incident intensity. Solid-state devices are fundamentally limited by the damage threshold. Even if the metasurface is placed where the laser spot diameter is 100 mm (approaching current physical and economic manufacturing limits[42]), the resulting intensity for a 5 PW laser is still approaching $10^{14}$ W cm$^{-2}$. This intensity already exceeds the damage threshold of state-of-the-art solid-state metasurfaces[43]. Our approach extends the metasurface functionality into the plasma regime, overcoming the intensity ceiling, albeit in a single-shot configuration. Yet single-shot operation is often indispensable for high-intensity laser applications. This necessity stems from the laser's extreme brightness and ultrashort pulse duration, coupled with the demand for ultrahigh spatial and temporal resolution in experiments. This applies particularly to ultrafast time-resolved studies, irreversible damage scenarios, and extreme condition transient phenomena. For instance, in PSMs-manipulated laser-pump, laser-plasma-driven X-ray-probe studies of ultrafast dynamics in materials, single-shot mode should be employed. This is because multiple pump pulses can alter the sample's state (e.g., through phase transitions or stress accumulation) or cause damage, thereby obscuring the intrinsic dynamics. Similarly, a single laser-plasma-driven detection pulse is required to avoid signal averaging. Many existing high-power devices, such as plasma mirrors (used for improving laser contrast), are inherently single-use. Similar to plasma mirrors, which are laterally shifted to present a fresh surface for each laser shot (typically allowing for several tens of shots per substrate), PSMs can likewise be operated this way. Thus the operational mode of PSMs aligns with other high-power plasma devices. Second, for a 200 TW laser and a damage threshold of $10^{13}$ W cm$^{-2}$, multi-shot operation requires metasurface diameters exceeding 50 mm, resulting in considerably high manufacturing costs. In addition, the massive volume of fabrication data poses significant processing challenges, particularly for non-periodic metasurface devices composed of nanoantennas with varying geometries and rotation angles. In contrast, PSMs enable the use of much smaller nanostructures, significantly reducing costs by orders of magnitude. Moreover, for cutting-edge scientific exploration, customized devices are often required. Large-scale, multi-shot metasurfaces are prohibitively expensive to fabricate and thus difficult to modify arbitrarily. In comparison, small-sized, low-cost PSM devices offer far greater flexibility for experimental use. Third, PSMs can serve a dual role: they function not only as modulators for ultraintense light but also simultaneously act as dynamic plasma-targets for high-intensity interactions. As demonstrated (Fig. S8 and S9), we employ phase-gradient PSMs as targets to achieve highly efficient relativistic surface plasmon coupling, resulting in exceptional particle acceleration. In such applications, the metasurfaces necessarily operate in the plasma state.

Recent advancements in laser technology have enabled the development of relativistically intense lasers with ultrahigh contrast[44]. Prepulse-to-pulse contrast ratios of up to $10^{-12}$ at 5 ps before the main pulse has been achieved at 100 TW class laser systems[45-47]. This technological progress now allows for ultraintense laser interactions with finely structured targets at the nanoscale. For instance, experiments have demonstrated that periodic grating structures with depths as small as 50 nm can withstand intensities of about $2.5 \times 10^{20}$ W cm$^{-2}$ [48]. Therefore, it is anticipated that PSMs can be implemented in the relativistically regime. For state-of-the-art multi-petawatt lasers, reported laser contrasts have reached $10^{-10}$ at 6 ps[49]. Even with a moderate contrast of only $10^{-6}$ for a 5 PW laser, placing metasurfaces in front of the laser focus, where the spot size is approximately 1 mm (a realistic size for metasurface fabrication), results in an intensity of around $10^{17}$ W cm$^{-2}$ on the metasurfaces. The prepulse, with an intensity of about $10^{11}$ W cm$^{-2}$ (below the threshold for dielectric-to-plasma phase transition), should not damage the metasurface structures before the arrival of the main pulse. Once the main pulse interacts with the PSM, it can be modulated and focused to extremely high intensities. Therefore, PSMs also offer significant potential to enable challenging ultraintense light modulations for multi-petawatt lasers.

The functionality of Pancharatnam-Berry metasurfaces arises from the interaction of light with electrons, governed by the geometric arrangement of subwavelength nanostructures. Excessive laser radiation pressure could deplete electrons from these nanostructures, impairing metasurfaces performance. To estimate an approximate upper limit for the applicable laser intensity, we employ a simplified one-dimensional model[50]. In this framework, the ponderomotive force induced by the laser $\left(v_e \times B_0/c \sim E_0\right)$ must remain below the maximum charge separation field that the nanostructure can sustain $\left(E_{\parallel} = 4\pi e n_{\mathrm{e}} d\right)$, where $v_{\mathrm{e}}$ is the electron velocity, $d$ is the nanostructure height, and $E_0$ and $B_0$ are the laser electric and magnetic field strengths, respectively. This gives the condition $a_0 < (n_{\mathrm{e}}/n_{\mathrm{c}})(2\pi d/\lambda_0)$. For a typical nanostructure with $n_{\mathrm{e}} \approx 400 n_{\mathrm{c}}$ and $d \approx 100$ nm, the maximum sustainable intensity is found to be on the order of $10^{23}$ W cm$^{-2}$. Notably, higher electron density or nanostructure height would allow the plasma metasurfaces to withstand a greater laser intensity.

Our study bridges metasurfaces and high-field science, two distinct yet highly active research fields. By exploring materials beyond conventional solids, metasurfaces composed of plasmas, which can withstand substantial light intensities, open up new opportunities for both metasurface and high-field research. Building on the foundation of our current proof-of-concept demonstration, the next step is to extend the experiments to higher laser intensities, particularly into the relativistic regime, to experimentally validate the robustness and functionality of PSMs under extreme conditions and to explore their potential applications in ultrafast, high-field photonics. This approach could lead to a new level of high-intensity light control through tailored design, even for state-of-the-art multi-petawatt lasers. The abilities to simultaneously generate three bright light beams with distinct polarization states, utilizing the photonic spin Hall effect, and intense stable-propagating optical vortices with integer topological charges, could lead to novel applications, such as two CP gamma ray vortices to enable single-shot measurements of strong-field quantum electrodynamic effects, including quantum radiation reaction, Breit-Wheeler pair production and even vacuum structures. Harnessing the powerful toolkit of PSMs is anticipated to have a profound impact on various related applications of high-field science, including high-intensity structured light, diverse particle acceleration techniques, and secondary radiation sources.

## Materials and methods

### Sample fabrication

The fabrication of the metasurfaces is carried out on a quartz substrate, involving the steps of patterning, deposition and lift-off. First, a PMMA495a4 positive resist layer is deposited by spin-coating on the quartz substrate. The sample is then exposed using electron beam lithography (EBL) and the pattern is revealed after the development process. Next, a 100-nm-thick gold layer is deposited onto the patterned substrate using electron beam evaporation deposition (EBD) method. Finally, the pattern is transferred onto the gold layer by the lift-off process, where the sample is immersed in hot acetone of 60° and cleaned by ultrasonic waves.

An individual gold cuboid has dimensions of 300 nm in length, 120 nm in width, and 100 nm in thickness. A unit cell is composed of eight cuboids with a cuboid-to-cuboid spacing of 400 nm in both $y$- and $z$-direction. The orientation of the cuboid is linearly rotated in the y-direction with a step size of π/8.

**Experimental setup**

The experiment is conducted using the 20 TW Ti:sapphire laser facility at Shanghai Jiao Tong University. This laser system can deliver 26 fs pulses centered at a wavelength of $\lambda_0 = 800$ nm, with an energy of up to 600 mJ at the exit of the CPA amplifier stages. The laser beam is focused onto a spot with a FWHM size of about 7.9 μm by a f/10 off-axis parabolic (OAP) mirror. The focal spot is imaged by a microscope objective and a CCD camera. The p-polarization laser beam is incident obliquely on the metasurfaces at an angle of about 35°.

For the photonic spin Hall effect, the three reflected beams are collected by a f/4 OAP mirror followed by three f/8 OAP mirrors and measured using three CCD beam profilers (Ophir SP392U). To analyse the polarization states of the two circularly polarized beams, two quarter-wave plates are individually inserted along the anomalously reflected trajectories, followed by two linear polarizers. The polarization state of the linearly polarized beam by normal specular reflection is detected using a single linear polarizer.

To measure the lifetime of the plasma-state metasurface structures, a small portion of the main laser pulse is split as a probe beam. Both incident beams are first aligned using a servo mirror system and then focused onto the target by the same OAP mirror at different incident angles, and thus six distinct reflected beams are observed. The pump and probe beams are carefully adjusted to achieve spatial and temporal overlap. The reflected intensities of the probe beam are recorded as the time delay between the pump and probe beams is varied. A fresh target is provided for each pump-probe measurement.

To generate optical vortices, the incident laser beam is first converted to circular polarization using a quarter-wave plate. The reflected vortex light is then measured by a CCD beam profiler. Prior to detection, the PB-phase-unaffected Gaussian pulse (circularly polarized with the opposite handedness to the vortex beam) is filtered out using a combination of a quarter-wave plate and a linear polarizer.

**Numerical simulation**

The 3D PIC simulations are performed using the fully electromagnetic relativistic PIC code VLPL (Virtual Laser Plasma Lab), which self-consistently solves the Maxwell equations for the electromagnetic fields and the relativistic equations of motion for the collisionless plasma macroparticles. The laser pulses propagate along the *x*-axis, incident normally onto the target surface. The laser wavelength is $\lambda_0 = 800$ nm, and the normalized laser amplitude $a_0 = eE_0/m_{\mathrm{e}}\omega_0 c$ is set to 1.5 ($a_0 > 1$ corresponds to the relativistic intensity regime), where $E_0$ is the laser amplitude, $\omega_0$ is the laser angular frequency, $e$ is elementary charge, $m_{\mathrm{e}}$ is electron mass, and $c$ is the light speed in vacuum. This corresponds to a laser intensity of $2 \times 10^{18}$ W cm$^{-2}$ for linearly polarized lasers.

For the photonic spin Hall effect simulations, the simulation box size is $x \times y \times z = 25\lambda_0 \times 24\lambda_0 \times 0.5\lambda_0$. The grid step size is $\Delta x \times \Delta y \times \Delta z = 0.01\lambda_0 \times 0.01\lambda_0 \times 0.0125\lambda_0$. Absorption boundary conditions are used in the *x*- and *y*-directions, while periodic boundary conditions are applied in the *z*-direction for both fields and particles. The temporal laser profile follows a Gaussian envelop $\exp(-t^2/2\tau_0^2)$ with $\tau_0 = 2.12T_0$. The transverse laser profile is Gaussian in the y-direction, $\exp(-y^2/2y_0^2)$, with $y_0 = 2.5\lambda_0$, and a plane wave in the *z*-direction. When simulating the light response of an individual cuboid meta-atom, the simulation box size is reduced to $x \times y \times z = 25\lambda_0 \times 0.5\lambda_0 \times 0.5\lambda_0$. The laser is a plane wave in both the *y*- and *z*-directions, and periodic boundary conditions are applied for both fields and particles. The other parameters remain unchanged from those used in the photonic spin Hall effect simulations. For the simulations of optical vortex generation, the simulation box size is $x \times y \times z = 20\lambda_0 \times 20\lambda_0 \times 20\lambda_0$. The grid step size is $\Delta x \times \Delta y \times \Delta z = 0.01\lambda_0 \times 0.02\lambda_0 \times 0.02\lambda_0$. Absorption boundary condition is used in all three directions for both fields and particles.

The targets are composed of full ionized plasmas with an electron density of $n_{\mathrm{e}} = 400n_{\mathrm{c}}$. The front surface of the target is located at $x = 20\lambda_0$. Due to the ultrashort timescale, the ions are assumed to be immobile. Each cell is filled with 16 macroparticles.The plasma metasurface geometry is chosen to match that of the experimental samples. Individual cuboid meta-atoms measure $0.375\lambda_0$ (300 nm) in length, $0.15\lambda_0$ (120 nm) in width, and $0.2\lambda_0$ (160 nm) in thickness. A unit cell comprises eight cuboids spaced $0.5\,\lambda_0$ (400 nm) apart in the *y*-direction. The metasurfaces consist of four unit cells arranged in the *y*-direction with a periodicity of $4\lambda_0$ (3200 nm). The orientation of the cuboids is linearly rotated with a step size of π/8.

## Acknowledgements

The authors thank Weiren Zhu for helpful discussions and Guangzhou Geng, Shuo Du, and Zhenfei Li for providing additional samples. This work is supported in part by the National Key

R&D Program of China (no. 2021YFA1601700), Fundamental and Interdisciplinary Disciplines Breakthrough Plan of the Ministry of Education of China (JYB2025XDXM204), the National Natural Science Foundation of China (no. 12175157, 12074251, 12225505, U21A20140, 12335016, and W2412039), the Beijing Natural Science Foundation (no. JQ24028), and the strategic Priority Research Program of the Chinese Academy of Sciences grant (no. XDA25010500 and XDA25010100). S.Z. acknowledges the financial support from the New Cornerstone Science Foundation, Hong Kong Research Grant Council (no. AoE/P-502/20, STG3/E-704/23-N, and 17309021), and Guangdong Provincial Quantum Science Strategic Initiative (no. GDZX2204004 and GDZX2304001). The authors would also like to acknowledge the Al for Science Program, Shanghai Municipal Commission of Economy and Informatization (Grand No. 2025-GZL-RGZN-BTBX-02029) and the sponsorship from the Yangyang Development Fund.

## Author contributions

Z.-Y.C. conceived the idea. W.Y. administrated the project. S.Z. and J.Z. supervised the project. W.Y., H.X., J.J., Z.-Y.C., S.C., M.W., M.M., R.L., and M.L. performed the experiments with assistances from G.L., W.Z., H.M., Z.Z., Z.Y., J.G., F.L., B.L., M.C., and L.C. H.L., Y.C., Y.Z., and Y.W. fabricated the samples and carried out the SEM measurements. Z.-Y.C. performed the simulations with support from F.Y. Z.-Y.C. drafted the manuscript with input from W.Y., L.H., S.Z., and J.Z. and final approval from all authors.

## Data Availability

The data that support the findings of this study are available from the corresponding authors upon request.

## Conflict of interest

The authors declare no competing interests.